**Extraction of a structural short-range order descriptor from nanobeam electron diffraction patterns using a transfer learning approach**

Junjie Wu [1], Timothy J. Rupert [1,2,*]

[1] Department of Materials Science and Engineering, Johns Hopkins University, Baltimore, MD 21218, USA

[2] Hopkins Extreme Materials Institute, Johns Hopkins University, Baltimore, MD 21218, USA

[*] Corresponding Author: tim.rupert@jhu.edu

## Abstract

Amorphous solids exhibit structural short-range order despite lacking long-range crystalline order, with this structural descriptor found to be important for determining mechanical properties. Nanobeam electron diffraction offers a potential route for experimental characterization of structural short-range order, yet efforts to date have been primarily qualitative in nature. In this work, machine learning approaches based on transfer learning are used to enable quantitative analysis of nanobeam electron diffraction data from amorphous solids. A ResNet-18 model is trained on simulated diffraction patterns taken from different locations within simulated metallic glasses and amorphous grain boundary complexions in the Cu-Zr alloy system that were created with hybrid molecular dynamics and Monte Carlo simulations. The disorder parameter is found to be a superior target structural descriptor compared to traditional Voronoi indices for this task. The model achieves a low validation mean absolute error across diffraction patterns corresponding to different interaction volumes, demonstrating excellent performance and potential transferability. Testing was performed using other simulated nanobeam electron diffraction data as well as experimental nanobeam electron diffraction patterns, showing that the model can reliably capture spatial variations in local structural state. As a whole, this framework is able to overcome the challenges in the quantitative experimental characterization of structural short-range order, enabling improved characterization of amorphous solids and the exploration of structure-property relationships.

## 1. Introduction

Amorphous solids are materials characterized by a lack of the long-range order typically seen in crystalline materials. Despite the absence of structural periodicity, amorphous solids exhibit unique properties due to their underlying local structural arrangement (e.g., see Ref. [1]), which can be referred to as structural short-range order (SSRO). One example of amorphous solids are metallic glasses, which have attracted significant research interest since they were first reported in 1960 [2]. Compared to their crystalline counterparts, metallic glasses are notable for their mechanical properties, such as high strength [3–8] and elastic limit [9–12]. At the same time, their practical applications are often restricted by their limited toughness [13,14], tensile ductility [15,16], and fatigue endurance [17,18]. The interplay between these properties and SSRO has been shown to be critical. First, SSRO is closely connected to the yielding behavior and shear band propagation in metallic glasses [19,20]. Atomic motifs such as full icosahedra are known to exhibit low potential energy, which enhances the strength and stability of metallic glasses [21]. Meanwhile, a higher concentration of densely packed icosahedra motifs suppresses atomic mobility, thereby reducing ductility [22]. Furthermore, the spatial heterogeneity of SSRO creates regions with varying atomic packing density, which can act either as crack barriers or as stress concentrators. In contrary to full icosahedra, geometrically unfavored motifs (GUMs), associated with higher potential energies and lower atomic packing fractions, can facilitate local atomic rearrangement under stress [23,24]. While such motifs contribute to a degree of plasticity, they also promote strain localization and propagation of shear bands, ultimately decreasing the toughness and fracture resistance of metallic glasses [25–27]. These relationships between SSRO and mechanical properties show the critical role of SSRO in the design of practical amorphous solids.

Amorphous grain boundary complexions, where amorphous intergranular films are formed through solute segregation between two crystals, are another example of a promising amorphous solid. For example, amorphous grain boundary complexions in nanocrystalline alloys have been shown to improve toughness and ductility by acting as damage tolerant interfacial features [28,29]. Mechanical testing of amorphous-crystalline nanocomposites demonstrates a similar simultaneous improvement of both strength and ductility [30,31]. Amorphous grain boundary complexions can act as strong dislocation sinks [32], reducing local stress concentration and resist crack nucleation at the interface [33]. Even subtle patterning of SSRO can be important for the mechanical behavior of amorphous complexions. For example, the asymmetry of SSRO between the amorphous-crystalline transition regions (ACTRs) on either side of the complexion were found to dramatically affect damage tolerance [34]. Similarly, amorphous grain boundary and interphase complexions were shown to provide superior radiation damage tolerance in Cu-Ta alloys compared to ordered boundaries, which was attributed to a heterogeneous distribution of excess atomic volume that facilitated more efficient defect recombination and annihilation [35].

While the importance of SSRO is clear, efforts to study such features are often restricted to atomistic simulations. Case in point, all of the literature examples discussed above relied on molecular dynamics (MD) simulations to probe the internal structure of amorphous solids. In an atomistic simulation, one can directly discuss SSRO because all atomic positions are known and local atomic packing can be directly observed and measured. In contrast, experimental study of SSRO remains challenging due to inherent difficulties in directly observing and quantifying nanoscale structural arrangements. High resolution transmission electron microscopy (TEM) has been used to analyze chemical short-range order of crystalline alloys [36–38], as the resolution of a TEM is fine enough to visualize sub-nanometer features, yet the lack of translational symmetry in amorphous solids and the fact that TEM images are two-dimensional projections from a three-dimensional volume means that SSRO cannot be directly extracted from regular imaging. Complementary techniques such as atomic pair distribution function analysis can provide structural information on a global scale, yet cannot resolve local SSRO or its patterning that builds the overall amorphous network [39].

Nanobeam electron diffraction (NBED) is promising for measuring SSRO and uses a small, focused electron beam to probe local atomic arrangements. For example, Hirata et al. [40,41] identified geometric frustrated motifs in metallic glasses by comparing experimental NBED patterns with simulated diffraction data and finding qualitative similarities. As the beam size increases and large atomic volumes are probed, the diffraction pattern acquired gradually converges toward the halo rings seen in regular electron diffraction [40], meaning a small beam is required for probing SSRO. If information on the spatial distribution of local packing is sought, NBED patterns can be taken at every position across a two-dimensional grid in the sample through a technique known as four-dimensional scanning transmission electron microscopy (4D-STEM) [42]. NBED patterns and, by extension, 4D-STEM datasets are rich in information, yet are not easily converted into the same metrics one can take from atomistic simulations. Recent efforts have explored the use machine learning (ML) techniques to analyze NBED data from both crystalline and amorphous solids [43–46]. For example, Kang et al. [47] performed ML-assisted pair distribution function analysis to categorize a 4D-STEM dataset into local structural types. However, classification models do not distinguish between the different types of SSRO one can find in amorphous regions. Bruefach et al. [48] used unsupervised non-negative matrix factorization to extract features from raw diffraction patterns, which does not require prior knowledge but requires image calibration, yet the extracted features require additional analysis to interpret. Consequently, there remains a need for a regression model capable of directly extracting a quantitative metric for SSRO from NBED patterns. Such a model would effectively bridge the gap between computational and experimental investigations of SSRO, and enable large-scale spatial mapping of structural order/disorder within amorphous solids.

In this work, an ML framework is designed to predict local SSRO from NBED patterns in the form of a local disorder parameter. By selecting a disorder parameter based on bond-orientation order [49], the model can capture the gradient of local SSRO in amorphous solids in terms of a metric that is commonly used in atomistic simulations. To ensure the robustness of our neural network in identifying diffraction patterns across varying disorder levels, the model is tested against a number of amorphous-crystalline configurations including amorphous grain boundary complexions and amorphous-crystalline composites. Additional validation is performed on experimental NBED patterns from the literature. As a whole, this study establishes a reliable approach for analyzing and interpreting complex diffraction patterns from NBED of amorphous solids.

## 2. Methods

### *2.1. Computational models*

To generate labeled diffraction patterns for supervised learning, we first created a set of atomistic Cu-Zr structures spanning the order-disorder spectrum from crystalline to fully amorphous. Hybrid molecular dynamics/Monte Carlo (MD/MC) simulations were performed using the Large-scale Atomic/Molecular Massively Parallel Simulator (LAMMPS) software [50] with an integration time step of 1 fs for all MD runs. Atomic interactions were modeled using the semi-empirical potential developed by Mendelev et al. [51] for the vitrification of Cu-Zr glasses.

Amorphous grain boundary complexions were simulated and used as the primary source of training structures because they exhibit pronounced SSRO gradients that transition from crystalline order in the grain interiors to amorphous structure within the complexion interior, passing through an intermediate structure along the way [52]. The simulation starts with a Cu-Cu bicrystal, in which the two face-centered cubic (FCC) grains are oriented in $[110]$ and $[00\bar{1}]$ along the *X*-direction, respectively, and both in $[1\bar{1}0]$ along the *Z*-direction. The bicrystal was heated to and maintained at 1000 K for 100,000 steps using isothermal-isobaric (NPT) time integration, before being doped with 2 at.% Zr through hybrid MD/MC simulations using the variance constrained semi-grand canonical (VG-SGC) ensemble. MC swaps were attempted every 100 MD steps, and the chemical potential was adjusted accordingly to control Zr concentration every 1,000 MD steps, with a total of 3,000,000 MD steps performed. The structure after minimization were visualized with OVITO software [53], as shown in Figure 1(a) and (b). The simulation cell measured 104.07 nm in *X*-direction, 6.28 nm in *Y*-direction and 41.71 nm in *Z*-direction. Common neighbor analysis (CNA) was used to identify crystalline and defect regions. A zoomed view of the grain boundary region is shown in Figures 1(c) and (d), where a fully amorphous structure that is heavily doped with Zr and approximately 8.7 nm thick along the *X*-direction is observed. In Figure 1(c), a few atoms appear yellow, signifying the identification of an icosahedral structure. Figures 1(e)-(g) shows the radial distribution functions (RDFs) of different regions in the grain boundary complexions. The amorphous regions exhibit a strong first-neighbor peak at ~2.7Å and lack the long-range periodic peak structure characteristic of the crystalline region, further confirming that the structure is amorphous.

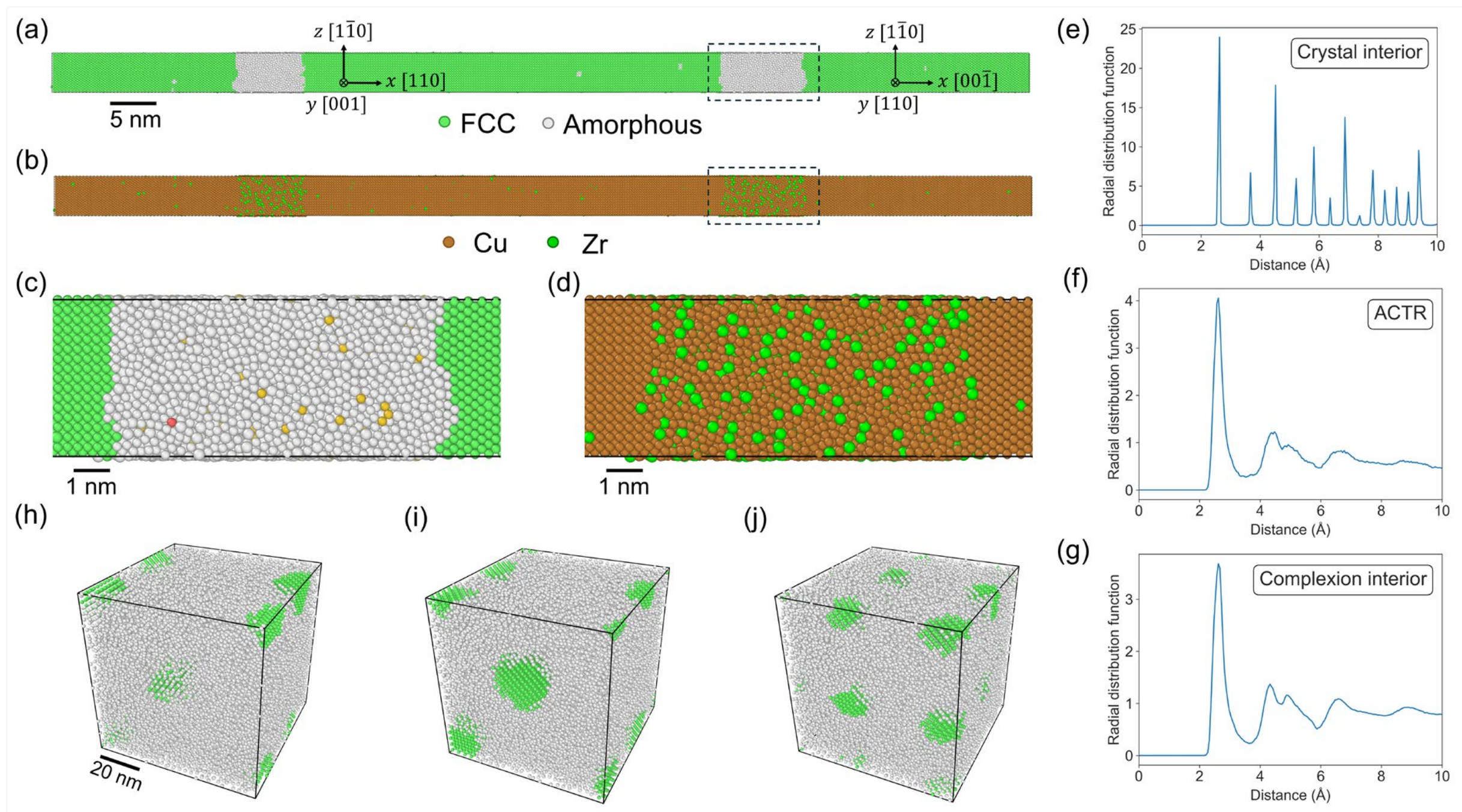

**Figure 1. (a, b) Front view of the equilibrated amorphous grain boundary complexion sample. (c, d) Enlarged views of the amorphous grain boundary complexions formed between the crystalline Cu grains. Radial distribution functions for the (e) crystal interior, (f) amorphous-crystalline transition region (ACTR), and (c) complexion interior of the amorphous grain boundary complexions. Metallic glasses containing nanocrystals with radii of (h) 4× the lattice parameter of Cu and varied orientations, (i) 4× the lattice parameter of Cu and aligned orientations, and (j) 3× the lattice parameter of Cu and aligned orientations. (a, c, h-j) are colored by CNA while (b, d) are colored by particle type, with legends provided in the figure.**

Amorphous-crystalline composite models were created to test the transferability of the model. Begin with bulk Cu crystals as the starting structure, spherical regions with radii of 4× or 3× the lattice parameter of Cu were kept frozen, while the remainder of the sample was doped by converting 36% of the atoms to Zr and melted by heating the system to 2000 K. The system was held at this temperature and everything except the small frozen crystalline regions became liquid. The samples were then cooled to 700 K with a quenching rate of 100 K/ns. Next, the nanocrystals were released from their frozen state, and the system was further equilibrated at 700 K for 2 ns. Finally, the entire system was cooled to 300 K at the same rate with a quenching rate of 100 K/ns. An additional sample was created following the same procedure, with the nanocrystals rotated with respect to one another. The final amorphous-crystalline nanocomposite samples are shown in Figures 1(h)-(j), where small crystalline regions (green) are seen scattered throughout an amorphous matrix (white). The RDFs for the amorphous matrices of those composites also show a first-neighbor peak at ~2.7 Å and confirm their amorphous structures, but are not shown here for brevity.

### *2.2. SSRO analysis*

The quenched system was analyzed with multiple techniques to evaluate SSRO. One of these is Voronoi tessellation, which partitions space into polyhedra by assigning each point to the atom to which it is closest. The resulting Voronoi polyhedron around an atom provides a geometry-based description of its local atomic environment. A common way to describe each polyhedron is by a Voronoi index in the form $<n_3, n_4, n_5, n_6>$, where $n_i$ represents the number of faces in the polyhedron with $i$ edges [54]. Representative examples of characteristic Voronoi polyhedra include FCC with index <0, 12, 0, 0> and icosahedron with index <0, 0, 12, 0>. In this work, Voronoi analysis is primarily used to track how SSRO varies spatially, which is useful for identifying regions such as the ACTR. However, Voronoi indices do not provide a unique or continuous measure of similarity between environments, and motifs with similar indices are not necessarily structurally similar. To mitigate the impact of small faces on the validity of Voronoi index, an edge length threshold of 0.6 Å was applied here.

Local SSRO can also be characterized by descriptors calculated from bond-orientation order [49,55,56]. Many atomistic studies of grain boundary structure employ a disorder parameter, $d_i$, which provides a measure of where a structure falls on the order-disorder spectrum. First, a normalized complex vector consisting of components $\hat{Y}_{lm} = \bar{Y}_{lm}/|\bar{Y}_{lm}|$ for each atom is computed, where $\bar{Y}_{lm}$ is the average of the spherical harmonics $Y_{lm}$. Next, the degree of similarity between an atom $i$ and its neighbors is given by:

$$s_{ij} = \sum_{m=-l}^{l} \hat{Y}_{lm}(i)\hat{Y}_{lm}^{*}(j) \quad (1)$$

The disorder parameter of atom $i$ is then given by:

$$d_i = 1 - \frac{|s_{ij}|}{N_i} \quad (2)$$

where $N_i$ denotes the number of neighboring atoms. $N_i$ was chosen to be 12 in this work because bulk Cu and Zr both have this coordination number in their crystalline states. The disorder parameter measures the average dissimilarity between an atom and its neighbors, such that a value of 0 corresponds to crystalline phases and a value of 1 indicates pure liquid that is fully disordered [56]. To provide comparison, a reduced bond-orientational order was also computed as follows:

$$1 - Q_l = 1 - \sqrt{\frac{4\pi}{2l+1}\sum_{m=-l}^{l} \bar{Y}_{lm}\bar{Y}_{lm}^{*}} \quad (3)$$

In this case, $l$ is set to 6. This reduced form is for the convenience of assessing the degree of disorder in a manner analogous to the disorder parameter.

### *2.3. Simulated diffraction and ML analysis*

Simulated diffraction patterns were generated using the *compute saed* command in LAMMPS [57], which computes k-space scattering intensity from the atomic coordinates within a user-defined real-space sampling region. Diffraction meshes of regions in three types of geometry were sampled (1) one-layer cluster, which consists of a center atom and its neighbors, (2) cylinder, and (3) ellipsoid. Detailed discussion of diffraction region geometries will be presented below in the Results and Discussion section. Cylindrical regions approximate a localized probe of collimated electrons passing through a specimen of finite thickness, where the cylinder height is analogous to the thickness of a TEM specimen and the cylinder diameter is analogous to the effective electron probe diameter. The ellipsoidal interaction volume shape represents an imperfect or defective version of the cylindrical shape. To mimic a 4D-STEM scan and to generate sufficient statistics across different regions, the centers of the sampled interaction volumes were placed on a regular grid with a step size of 2.5 Å in the scan direction. As this step size is smaller than the effective probe diameters considered in some cases, neighboring interaction volumes can overlap. This overlap is intentional, since it yields smoothly varying averages of $d_i$ across the ACTR while still producing distinct diffraction patterns due to changes in the local atomic configuration within the interaction volume. In an experimental setting, the appropriate choice of step size depends on the probe diameter and desired spatial resolution and, in some instances, very small probe sizes and step sizes have been reported. For example, Hirata et al. reported NBED probe sizes on the order of 0.36 nm in metallic glasses [41], while Gao et al. demonstrated 4D-STEM measurements with a probe diameter of about 0.6 Å and scan step sizes down to 0.2 Å in a 4D-STEM workflow [58]. Our simulation-based scan corresponds to a small-step and oversampled acquisition intended to probe sensitivity to gradual SSRO variations.

Unless otherwise noted, diffraction patterns were generated using a maximum scattering vector of $k_{\max}$ =1.0 Å$^{-1}$ with a reciprocal space mesh spacing of $\Delta k$ = 0.015 Å$^{-1}$. For the one-layer clusters, complete diffraction meshes were computed, from which diffraction patterns were sampled by slicing along different zone axes. For the cylinder and ellipsoid regions, diffraction meshes with zone axis parallel to the shape's major axis were computed and the diffraction patterns were then obtained by slicing the diffraction mesh perpendicular to the zone axis. The raw intensity values were then log-transformed and normalized to the [0, 1] range before used as ML inputs. The resulting images were cropped or resampled to 133×133 pixels, corresponding to a k-space field-of-view set by $k_{\max}$ and $\Delta k$. In the context of an experiment, $k_{max}$ provides the effective outer collection angle (i.e., the highest spatial frequency included, analogous to a camera length dependent k-range), while $\Delta k$ sets the reciprocal space sampling resolution.

To expedite the training process of a highly accurate model, transfer learning technique was used. A pre-trained ResNet-18 [59], capable of classifying 1000 classes on ImageNet, was further trained for 400 epochs using the Adam [60] optimizer with a learning rate of 0.001 and mean absolute error (MAE) loss. A learning rate scheduler was applied to reduce the learning rate by half on plateau. Additional details and discussion of the model will be presented below in the Results and Discussion section.

## 3. Results and Discussion

To begin, it is important to establish an understanding of the ground truth for SSRO within the modeled materials. Here, an amorphous grain boundary complexion is used as a representative example, since it contains both ordered and disordered regions. Figure 2 shows the spatial variation of the three SSRO descriptors across the complexion sample: (1) the distribution of representative Voronoi polyhedra, (2) the disorder parameter $d_i$, and (3) the reduced bond-orientational order $(1 - Q_6)$. Starting with the Voronoi polyhedra, which have been extensively used in the computational study of SSRO in various types of amorphous solids [19,61–65], Cu is chosen as the central atom since it is the dominant atom present while the <0, 10, 2, 0> and <0, 2, 10, 0> motifs are specifically shown since they are close to the well-defined FCC and icosahedral structures. The <0, 10, 2, 0> polyhedra predominantly appear as two peaks near the ACTRs, while the <0, 2, 10, 0> polyhedra are present only in the complexion interior. Figure 3 shows a more detailed description of the amorphous complexion with the eight most frequent non-FCC Voronoi motifs shown. The ACTR is primarily filled with the <0, 10, 2, 0> and <0, 8, 4, 0> polyhedra, dominating the regions where *X*-position ranges from around 185 to 193 Å and from 277 to 286 Å. The ACTR regions appear slightly less than 1 nm in width in this figure, although this is a superposition of the actual width (~5 Å) plus some interfacial roughness that is added in such an averaged profile. In contrast, these motifs fall to very low densities in the complexion interior, where the other six visualized motifs are more plentiful.

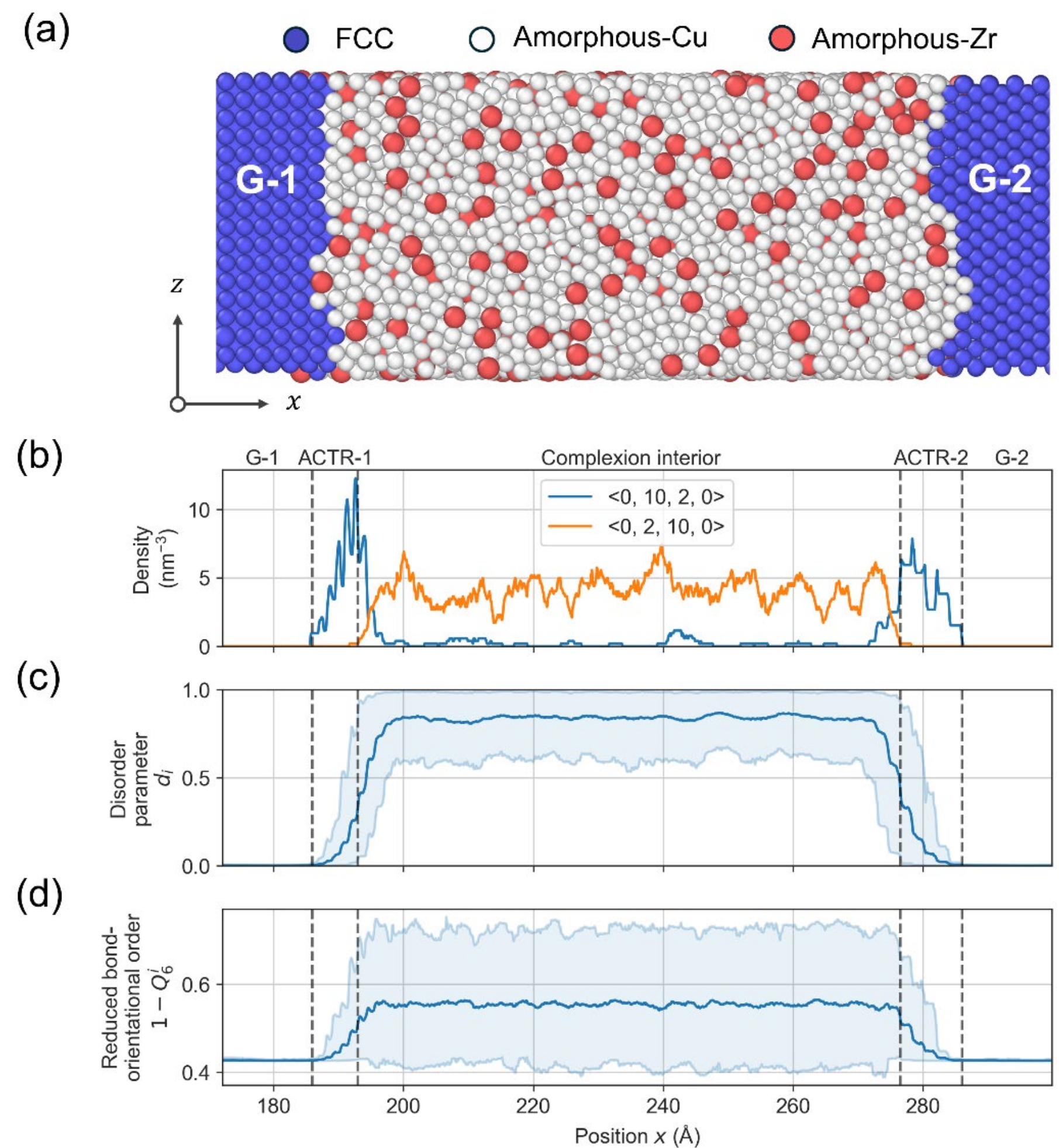

**Figure 2. (a) Visualization of an amorphous grain boundary complexion in Cu-Zr. Distributions of the (b) <0, 10, 2, 0> and <0, 2, 10, 0> Voronoi polyhedra with Cu as the central atom, (c) disorder parameter $d_i$, and (d) reduced bond-orientational order $1 - Q_6$. For (c) and (d), the center line and the upper and lower bounds represent, respectively, the mean, 95th percentile, and 5th percentile of all data points within a 2 Å sliding window centered at each position (±1 Å). The two grains are denoted G-1 and G-2, while the beginning and end of the ACTRs were chosen based on the positions where <0, 10, 2, 0> and <0, 2, 10, 0> polyhedra emerge, respectively.**

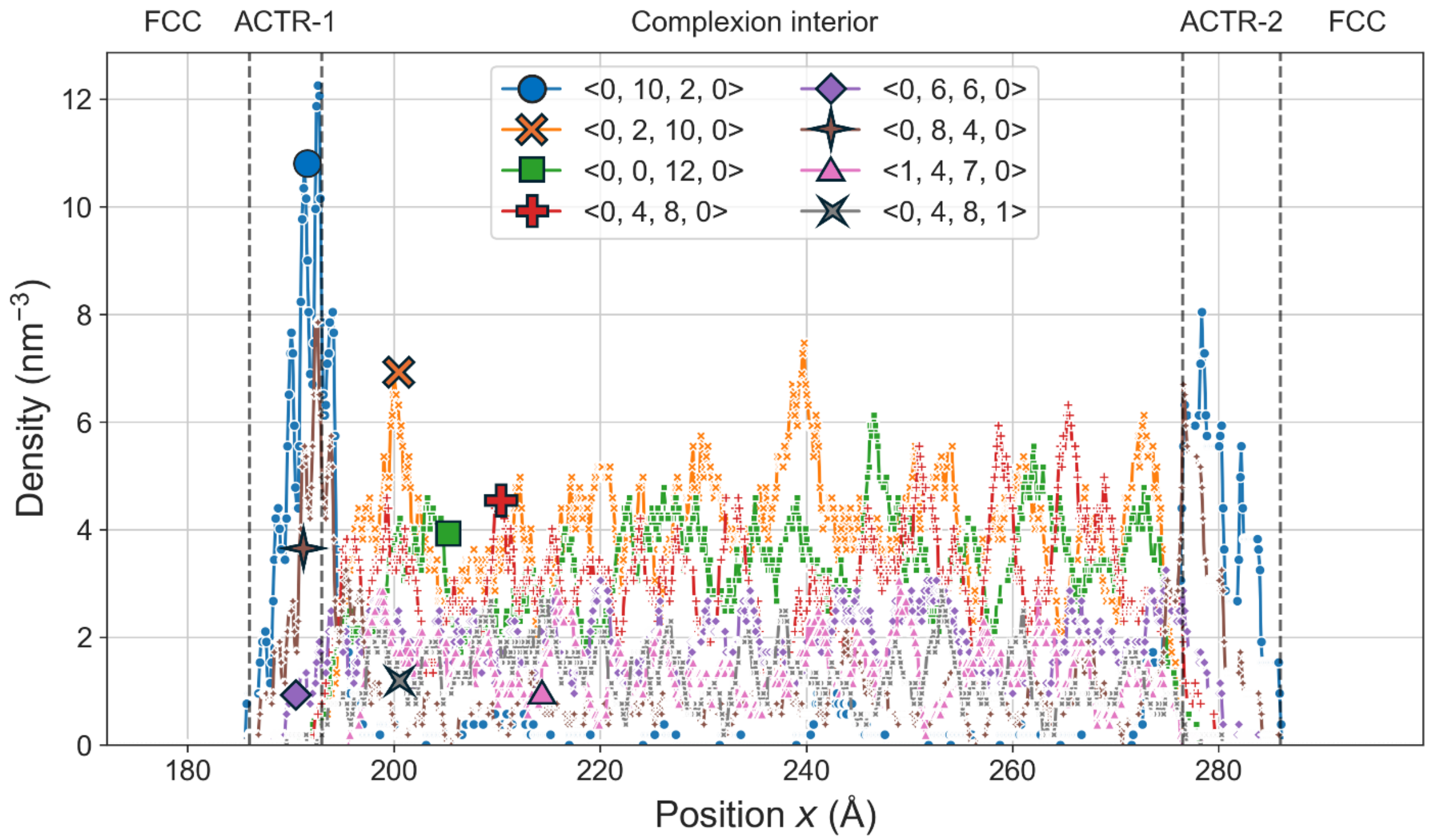

**Figure 3. Spatial density variations of the eight most frequent Voronoi motifs in the model (excluding FCC motifs) with Cu as the center atom along the thickness direction in the amorphous grain boundary complexion.**

Figure 2(b) and Figure 3 show that Voronoi polyhedra can be used to identify the ACTR and complexion interior, yet the use of Voronoi indices as an overall SSRO descriptor presents several challenges. First, a cluster of atoms can contain a number of distinct Voronoi indices due to the heterogeneity of local atomic arrangements in amorphous solids, meaning the assignment of a single value to the region is not possible. Second, there are a large number of Voronoi indices. Figure 3 shows the spatial density of the eight most frequent Voronoi polyhedra excluding FCC, yet this only represents 27.53% of the Cu atoms in the complexion. A total of 816 different motifs would need to be considered to fully describe the SSRO of the Cu atoms, making direct prediction of Voronoi indices impractical. If one where to include the Zr atoms, which have different baseline descriptions, even more motifs would need to be treated. This also highlights a common limitation of computational studies that is often overlooked. If Voronoi index is used to describe the local structure in an atomistic modeling study, the analysis often is limited to one atomic species because this enables an easy comparison. In an atomic scale model where one can simply ignore other species, this works, yet experimental signals will include all atoms. Classification models are possible yet lack the ability to describe specific structural states. However, amorphous regions with different degree of disorder are indistinguishable with such a scheme, meaning there is an oversimplification of atomic environments resulting in a loss of data granularity. Although it is possible to categorize Voronoi polyhedra into finer groups, such as the twelve building-block groups proposed by Weeks and Flores [66] or ordered and disordered classifications introduced by Garg and Rupert [67], details are still suppressed and an incomplete picture of SSRO would result. In addition, when mapping diffraction patterns to SSRO, the orientation dependence of local structure can create instability if discrete Voronoi indices are used as regression targets, because motifs that are geometrically similar can produce noticeably different diffraction patterns depending on distortion and viewing direction, and vice versa, distinct motifs can yield diffraction patterns with similar apparent symmetries. For example, a <0, 10, 2, 0> motif is structurally close to an ideal FCC <0, 12, 0, 0> motif, yet its diffraction pattern can differ significantly due to geometric frustration and local distortion. Conversely, different packing motifs, including distorted icosahedral environments, can display similar 2-fold or 6-fold symmetries when viewed along certain orientations [41].

To overcome these limitations, the disorder parameter $d_i$ (Figure 2(c)) and the reduced bond-orientational order parameter $1 - Q_6$ (Figure 2(d)) were explored as possible candidates for SSRO descriptor within an ML analysis environment. First and foremost, a successful descriptor must identify the key structural signatures of interest. The moving average (center line) of both $d_i$ and $1 - Q_6$ show similar trends across the amorphous grain boundary complexion. They begin at a nearly constant value within FCC regions, increase monotonically across the ACTR, and eventually reach a plateau within the complexion interior. However, the reduced bond-orientational order parameter ranges from approximately 0.40 to 0.75 in the complexion interior. Given that the reduced bond-orientational order parameter for FCC is 0.42, this descriptor provides a limited ability to distinguish between crystalline and amorphous regions, let alone comment on differences in SSRO. In contrast, the lowest data points from $d_i$ stay above 0.6 in the complexion interior, significantly differing from both the ACTR and the FCC grain interior. Disorder parameter thus provides an interpretable metric of disorder, ranging from 0 in the crystal to a mean value near 1 in the complexion interior. Additionally, for larger diffraction volumes, the gradients exhibited in the moving average $d_i$ suggest that the average $d_i$ of atomic clusters are still distinguishable between each other. It is important to note that the size of atomic cluster needs to be small enough to not wash out the signal, which will be discussed later as different diffraction volumes are probed.

Finally, for the disorder parameter to be a promising candidate as an SSRO descriptor, it must be able to distinguish the features of interest in an amorphous solid. As mentioned earlier, previous studies have highlighted the importance of certain groups of Voronoi motifs on mechanical properties. Figure 4(a) shows the distribution of $d_i$ for different Voronoi polyhedra categorizations: FCC, ordered, intermediate, and disordered. Here, the term "ordered" refers to motifs with $n_4 \geq 8$, $n_5 \leq 4$ and $n_5 = 0$, including <0, 10, 2, 0> and <0, 8, 4, 0> plus a few others, based on their (1) geometric similarity with FCC motifs (<0, 12, 0, 0>) and (2) spatial dominance in the ACTR [52]. Similarly, "disordered" motifs refer to Voronoi polyhedra with $n_4 \leq 2$, $n_5 \geq 10$ and $n = 0$, such as <0, 0, 12, 0> (icosahedron), <0, 2, 10, 0> (distorted icosahedron) and a few others that exist mostly in the disordered complexion interior. Starting from the FCC motifs, 98.9% of them have a $d_i$ lower than 0.01. The outliers are located near the amorphous-crystalline interface, as their $d_i$ shows the contextual information of their atomic environment that is missing in Voronoi analysis. While the atoms are part of the crystal, the region nearby is distorted. The median $d_i$ of ordered and disordered motifs are 0.290 and 0.915, respectively, clearly separating the two populations. By minimizing the total error rate when separating the ordered and disordered motif groups, an empirical threshold of $d_i$ = 0.768 was obtained for this dataset. Although the $d_i$ for ordered motifs exist over a broader range, 91.38% of the ordered amorphous motifs have a $d_i$ lower than 0.768, while 96.43% of the disordered amorphous motifs have a $d_i$ higher than 0.768.

The remaining, non-FCC polyhedra are referred to as "intermediate" motifs, which fill up the remaining space in amorphous structures. Among these are the motifs previously referred to as GUMs, which are geometrically unfavored for space tiling and energetically less stable [68,69]. There are of course hundreds of others that have no special name. The disorder parameter value for the intermediate structures, being neither completely ordered nor disordered, falls at intermediate values between those extreme yet important cases.

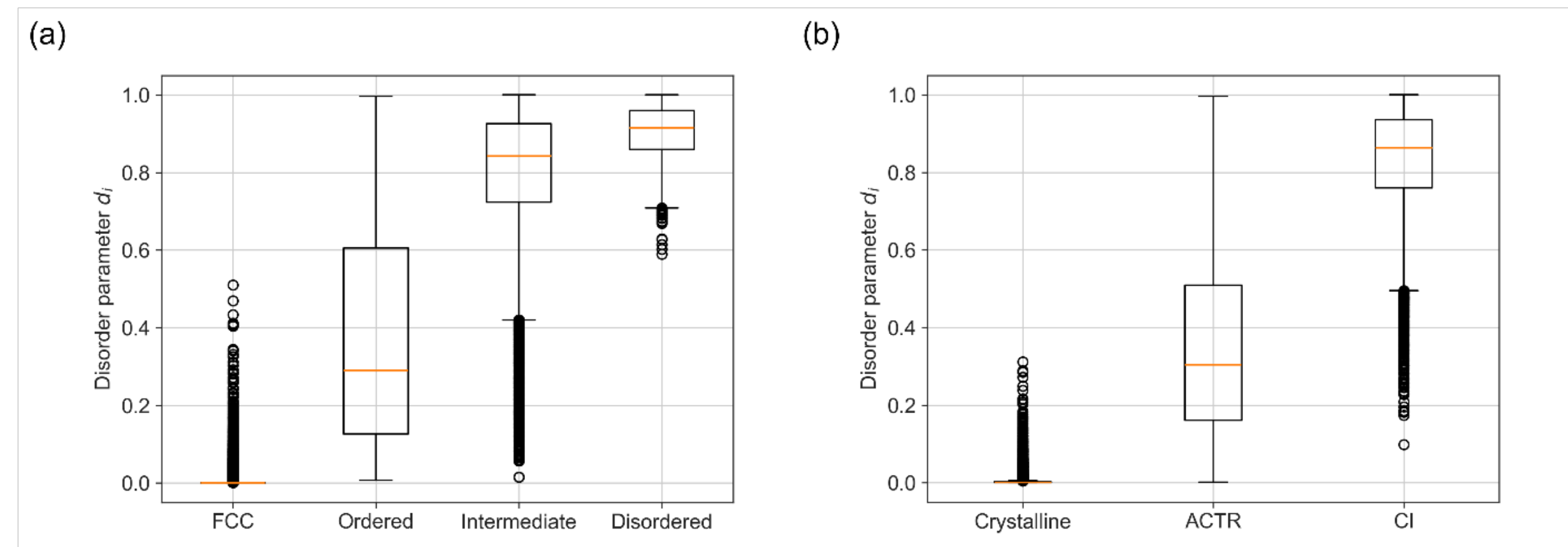

**Figure 4. Distribution of the disorder parameter $d_i$ according to (a) prior classification schemes from [67] and (b) the location within the amorphous grain boundary complexion. The boxes represent the first quartile to third quartile range, with a center line in orange representing the median. The whiskers extend 1.5 times the interquartile range from the boxes. Data points marked as circles denote the outliers lying outside of the end**

**of the whiskers. In (b), crystalline atoms were identified by CNA, while the criteria used to define the ACTR and complexion interior (CI) are described in Supplementary Note 2.**

The complexions themselves provide another way to further examine the rigor of the disorder parameter, as the ACTR and complexion interiors are noticeably different. Figure 4(b) presents the $d_i$ distribution in the crystalline grains, ACTR, and complexion interior. A distinct signal in $d_i$ can be found between the ACTR and the complexion interior, as their median values are 0.304 and 0.863, respectively. Using the empirical threshold of 0.768 discussed in the previous paragraph, 92.2% of polyhedra in the ACTR have a $d_i$ lower than the threshold and 73.5% of polyhedra in the complexion interior have a $d_i$ higher than the threshold. Some level of overlap here is to be expected, as a number of SSRO motifs are observed both in the ACTR and complexion interior. In other words, these regions are different yet do have features in common. Both the comparison of ordered and disordered packing motifs as well as the comparison of different interfacial environments show that the disorder parameter is capable of characterizing important variations in SSRO.

To build a model capable of predicting disorder parameter, the framework shown in Figure 5 was trained with simulated diffraction patterns from different positions across the amorphous grain boundary complexion. ResNet-18, a deep convolution neural network (CNN) featuring residual connections, was chosen in this task to because of its generalizability and fast convergence [59]. To accelerate model training, each iteration was initialized with weights pre-trained with ImageNet dataset and then these models continued being trained with simulated diffraction dataset. Each dataset consisted of simulated diffraction patterns with a size of 133×133 pixels that were computed from regions of three types of geometries (crystalline, ACTR, and complexion interior). The datasets for one-layer cluster, cylinder, and ellipsoid interaction volumes had 58593, 57720, and 137493 unique diffraction patterns, respectively. Datasets were shuffled and randomly split into training and validation subsets using a fraction of 0.8 and 0.2. Data augmentation was performed during the training phase by randomly applying transforms including rotation, random contrast, Gaussian noise, and random sharpness. While random rotations enforce rotation invariance, the other transforms introduce artificial noises to improve model's generalizability on experimental images. We note that these pro-processing steps could be revisited if one wanted to closely match the pre-processing of a specific 4D-STEM workflow, such as masking central transmitted beam.

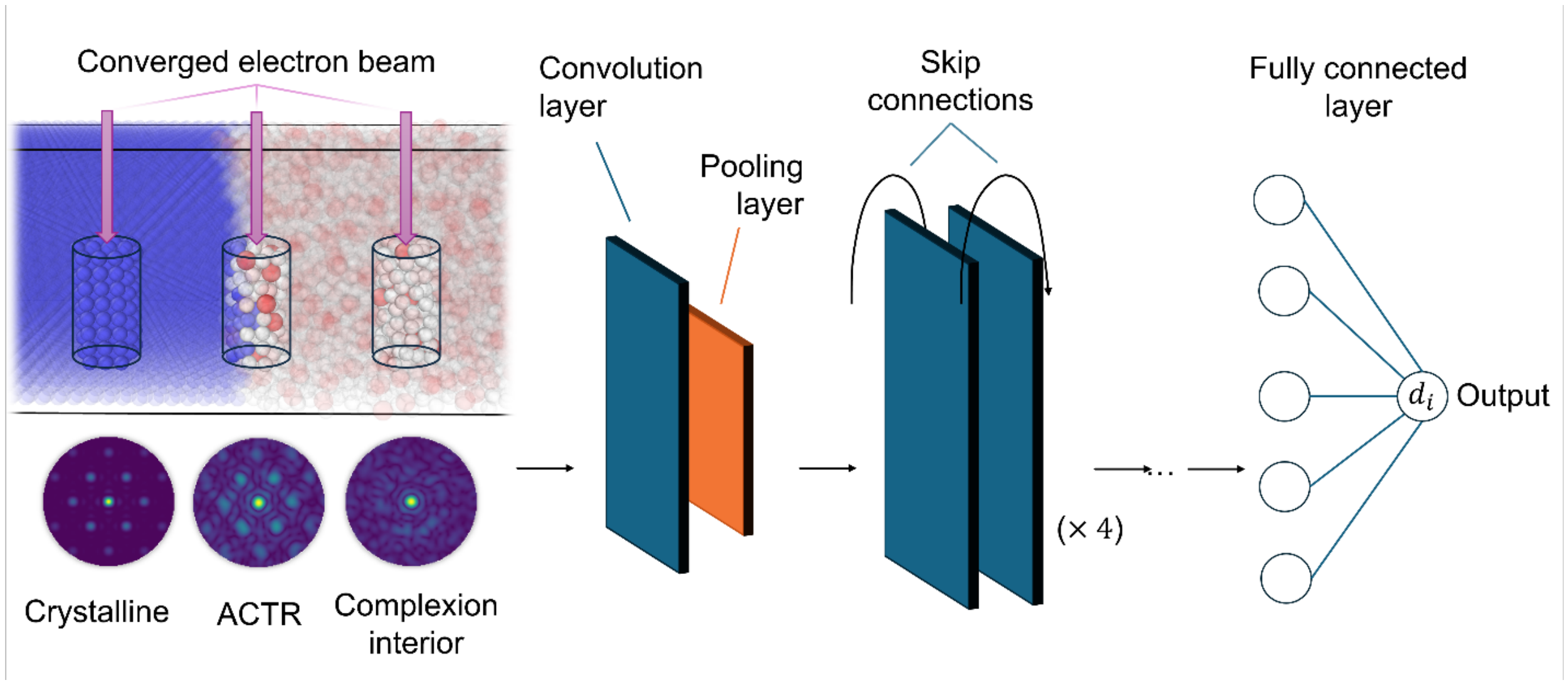

**Figure 5. Schematic illustration of machine learning analysis of an NBED dataset. Simulated diffraction patterns of regions sampled from different positions within the amorphous grain boundary complexion sample are used as inputs into a ResNet-18 model consisting of an initial convolution layer and pooling layer, 4 residual blocks with skip connections, and a fully connected output layer that transform high-dimensional data into predictions for disorder parameter $d_i$.**

To ensure the generalizability of the model and examine the limits on input diffraction patterns, it is necessary to connect the simulation volumes used to create the simulated diffraction patterns. Here, we explored three types of effective diffraction volumes: one-layer cluster, cylinder, and ellipsoid. The one-layer cluster contains only a central atom and its nearest neighbors, thus representing the smallest volume where one can describe SSRO. In contrast, the cylindrical and ellipsoidal shapes include collections of atoms within a finite interaction volume, in an attempt to be more representative of experimental NBED conditions where the recorded diffraction intensity arises from multiple atoms in the beam path. It is important to note that even though multiple atoms in a finite interaction volume are studied, SSRO is still the focus. Disorder parameter is calculated based on first-nearest neighbor packing, so it is inherently a measure of SSRO. The interaction volumes define which atoms contribute to the signal, yet we are still only extracting information about the SSRO distribution. Medium-range order, or the connectivity of atomic clusters over larger length scales, is not studied here. The cylindrical volume approximates a collimated electron beam traversing a thin TEM sample but introduces a sharp boundary, whereas the ellipsoidal volume provides a smoother boundary and mimics the higher weighting of scattering near the center of a focused probe. The ellipsoidal interaction volume can be thought of as a defective version of the perfect cylindrical interaction volume, and here is only used to provide insight into the effect of choosing a different interaction volume geometry. Future work could re-train the model using whatever interaction volume size and geometry most closely fits the details of a specific set of electron microscopy experiments. Figure 6 shows the performance of models trained with three different effective diffraction volumes, with coefficient of determination and root mean square error annotated on each. The models trained with cylindrical and ellipsoidal volumes have better performance (Figure 6(b-d)) than models trained with the one-layer clusters (Figure 6(a)), as confirmed by their coefficient of determination ($R^2$) and root mean square error (RMSE). In general though, all predictions are relatively robust. This is a positive outcome, as real NBED experiments will obtain a signal from more than a single atomic environment, with the cylindrical and ellipsoidal shapes representing simple models of typical electron beam interaction volumes. In contrast, the model for NBED of one-layer clusters is less physically realistic. Both cylindrical and ellipsoidal model perform well, with the model trained on cylindrical diffraction volumes slightly outperforming the one trained on ellipsoidal regions. Even for highly disordered regions where $d_i > 0.7$, the vast majority of data is still along the unity line. The interpretation of the magnitude of a predictive error generally lies in whether one can distinguish the behavior that is being studied. As such, we can define an acceptable error here as one that is much smaller than the variations in the physically meaningful environments we are studying. Relevant disorder parameter variations can be considered from Figure 4. From Figure 4 (a), the disorder parameter data without outliers for atoms in the "disordered" category spans ~0.25. Alternatively, from Figure 4(b), the disorder parameter data without outliers for atoms in the complexion interior spans ~0.5. As such, all of the RMSE value for the models with different geometries are much smaller than this, and therefore can distinguish different SSRO environments across amorphous complexions and other amorphous states. We also note that these are much smaller than the differences between the key structural states in the complexions (mean values of disorder parameter for the crystal is ~0, ACTR is ~0.3, and complexion interior is ~0.86 in Figure 4(b)), so the different regions can be identified based on disorder parameter as well. Notably, the disordered regions mostly reside in the complexion interior and the deviations from predicted values are small, meaning the model can still tell the different regions of the amorphous complexion apart. Due to better performance and closer connection to experimental probe geometry, we focus on the cylinder-trained model for the remainder of the manuscript.

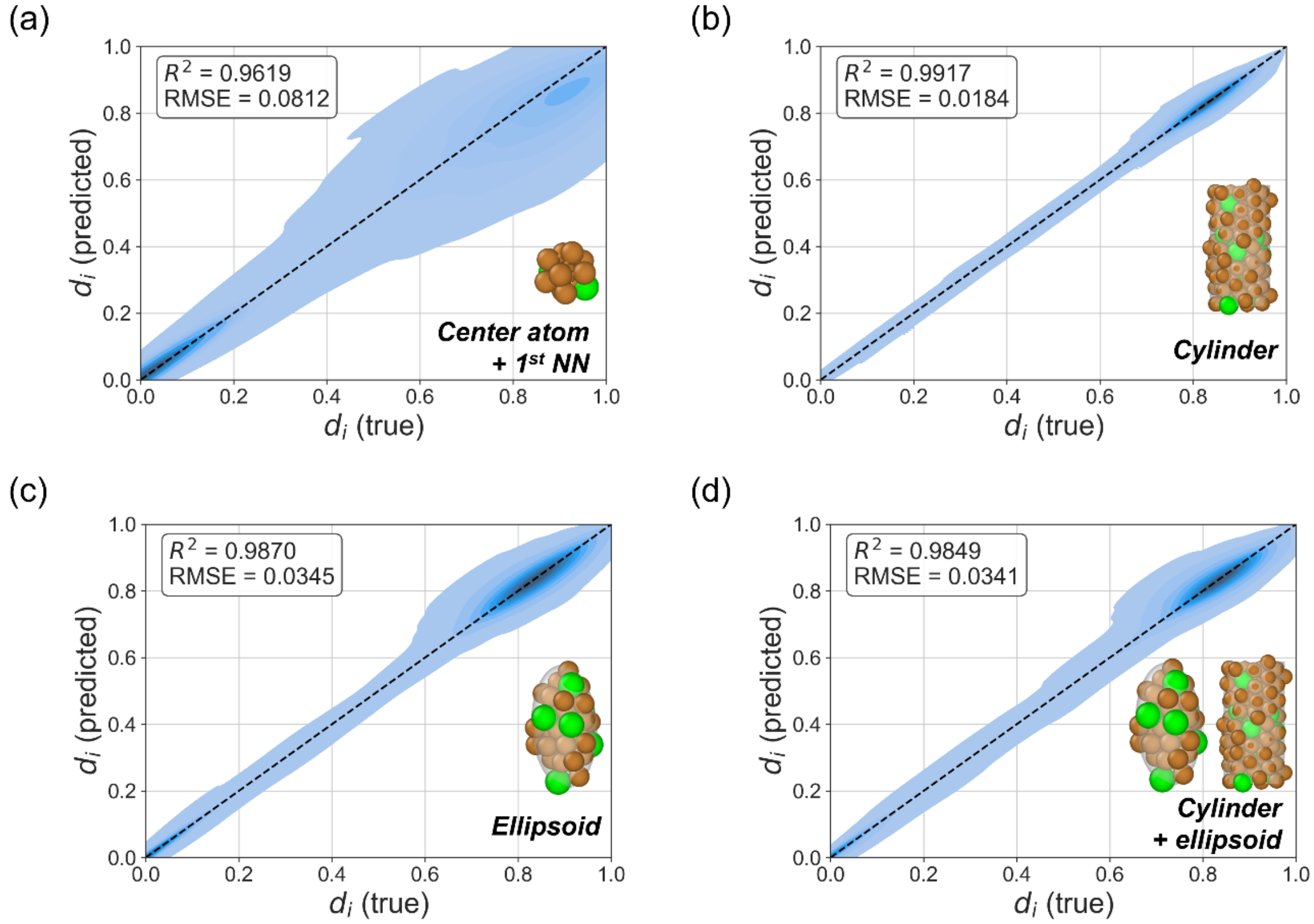

**Figure 6. Kernel density estimate plots comparing predicted and true values for the validation set of the ML model trained with simulated diffraction patterns from (a) one-layer clusters, (b) cylinders, (c) ellipsoids, and (d) a combined set of cylinders and ellipsoids. The black dashed diagonal indicates perfect agreement between prediction and ground truth. The coefficient of determination and root mean square error are reported on each subfigure. The outer contour corresponds to the kernel density threshold of 0.015.**

To further investigate the impact of effective diffraction volume geometry, this time in terms of experimental conditions such as beam size and sample thickness, the performance of the model trained with cylinders of different dimensions is plotted in Figure 7. We reiterate that these dimensions only describe the volume over which a collection of atomic disorder parameters is averaged, meaning that SSRO from local regions remains the focus. A base CNN model with three convolutional layers and two fully connected output layers (Figure 7(a)) was used for comparison with a model that used the pretrained ResNet-18 model (Figure 7(b)). Error is shown in RMSE, to be consistent with the other discussion of prediction errors in this study. For a given cylinder radius, intermediate values of cylinder height (i.e., sample thickness) were found to have the lowest loss. It is worth noting that while loss appears to be smaller for the highest height values as compared to the very small height values, this is primarily due to the diffraction patterns containing information from larger volumes and therefore converging toward a more averaged state that is easier to predict but also less useful. The observation that intermediate heights are best is consistent with experimental reports of NBED patterns with varying beam sizes. Small beam sizes were found to be difficult to interpret due to orientational dependence and spatial variation, while large beam size results in trivial halo rings [40]. However, ResNet-18 degrades much less severely in this small-volume regime, suggesting that it is better able to extract SSRO features when the effective probe size is small.

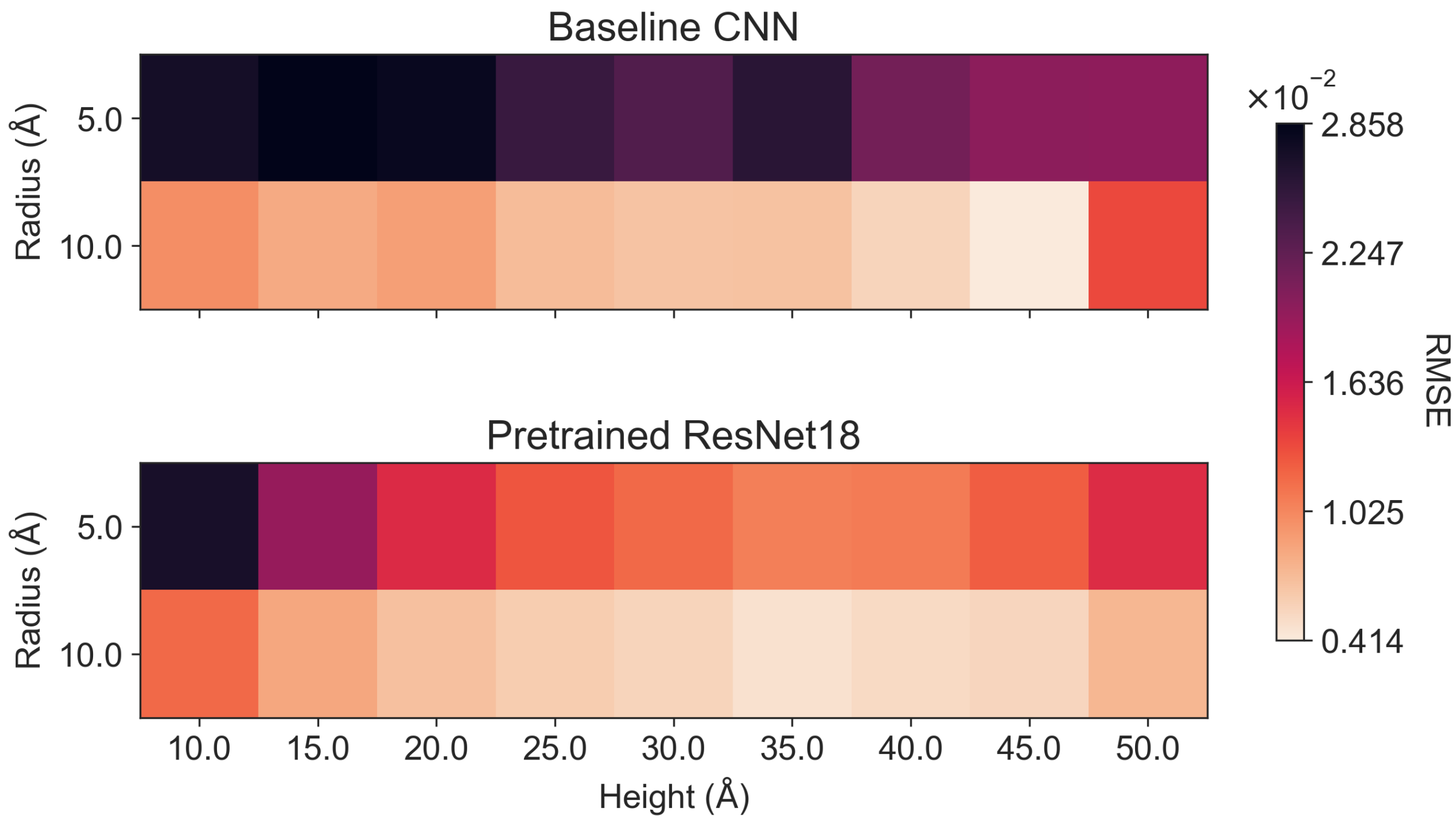

**Figure 7. Performance heatmap with respect to different cylindrical diffraction interaction volumes for (a) base CNN model and (b) the pretrained ResNet-18 model, with error shown as RMSE.**

The model was tested on simulated diffraction patterns not presented in the training dataset to understand its ability to predict unseen configurations. Figure 8 shows predictions of SSRO along two strips in amorphous grain boundary complexion samples where Figure 8(a) comes from the grain boundary used for training, thus providing a baseline expectation for how well the model can perform, while Figure 8(b) is from another grain boundary not used for training. The model was tested on simulated diffraction patterns not presented in the training dataset to understand its ability to predict unseen configurations. The predicted $d_i$ profile of unseen images remain quantitatively aligned with the ground truth, with an overall RMSE of 0.0313, as shown in Figure 8(b). When separated by region, the RMSE values for the crystalline volume, the ACTR, and the complexion interior are 0.0107, 0.0334, and 0.0385, respectively. Although the predicted profile does not perfectly reproduce ground truth values everywhere in the unseen data, the low error levels indicate that the model captures the local structural state with good accuracy. Even the locations with the largest deviations in the complexion interior are well within the expected values for the complexion interior motifs shown in Figure 4(b) and would be classified as "disordered" according to Figure 4(a). Within the complexion interior, $d_i$ reaches a plateau and the local fluctuations in $d_i$ are comparatively small relative to the strong gradient across the ACTR. The crystalline regions yield the lowest RMSE since there is little variation in the local structural packing in a crystal (nearly perfect FCC packing, with at worst small elastic strains). Notably, the strong quantitative agreement in the ACTR, where SSRO variations are largest and most structurally meaningful, supports the conclusion that the model generalizes across distinct amorphous complexion configurations and is capable of identifying subtle changes in SSRO. For example, within the ACTRs in Figure 4(b), it can also identify the subtle yet important asymmetry in SSRO between the two sides of the complexion. Specifically, ACTR-3 has a mean $d_i$ of 0.2767 and is predicted as 0.3035, while ACTR-4 has a higher mean $d_i$ of 0.3903 and is predicted as 0.3829. The ACTRs have been identified as particularly important in past works. For example, after creating amorphous grain boundary complexions with MD/MC simulations, Garg and Rupert [67] found that the SSRO in the ACTRs can be predicted by the incompatibility between the confining grains, described by the grain boundary strain. The analysis tool presented here will enable complementary experimental studies of these features.

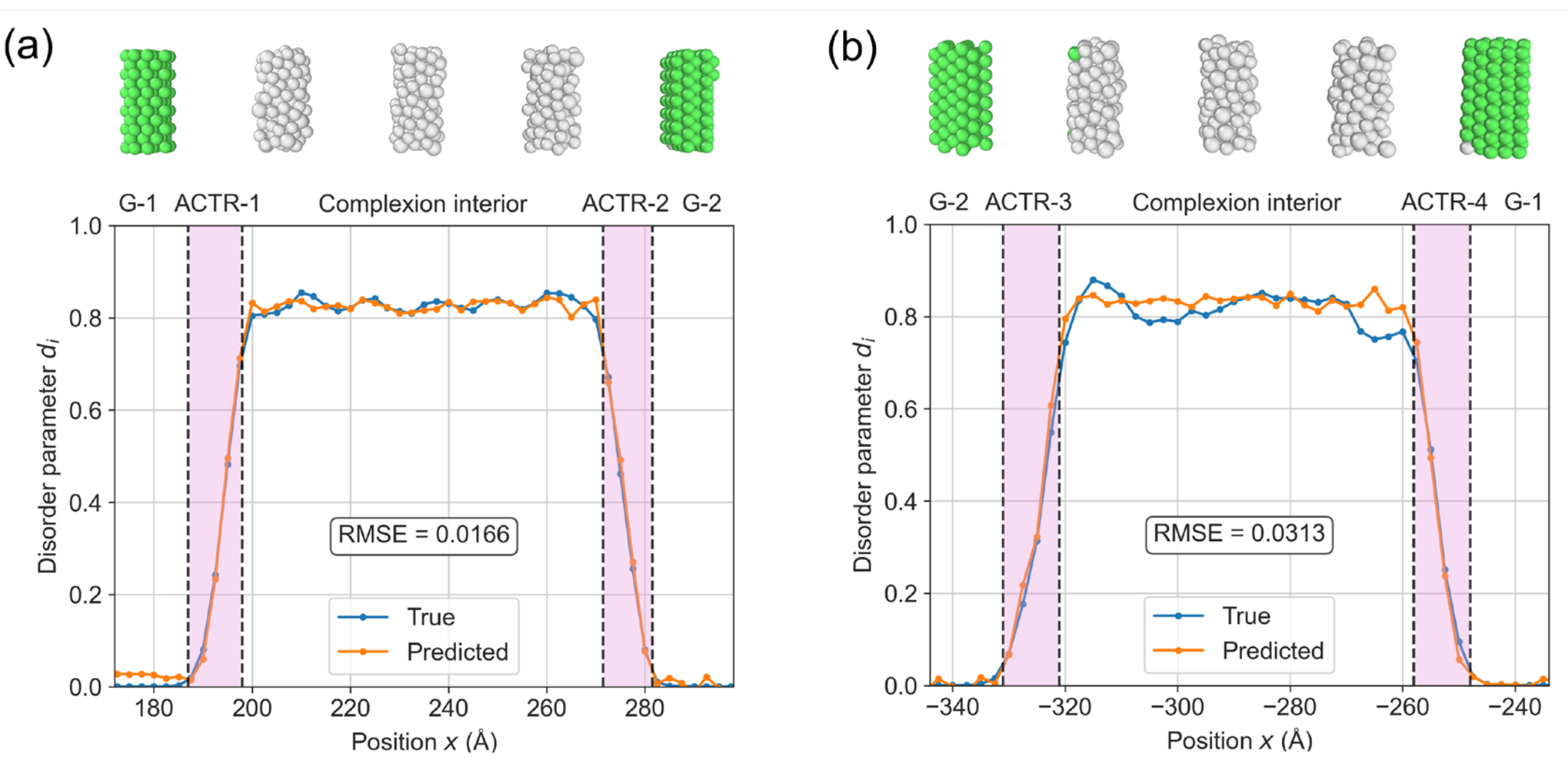

**Figure 8. Predicted and true $d_i$ profiles for simulated diffraction patterns taken from two strips across amorphous complexions (a) used for training and (b) not used for training. The regions highlighted in pink correspond to the ACTRs. For each data point, the value of the disorder parameter is the average of each atom-by-atom $d_i$ value in the region. All diffraction patterns were generated using a cylindrical interaction volume with radius $r$ = 5 Å and height $h$ = 20 Å. Cylinder centers were sampled with a 2.5 Å step size, leading to overlapping interaction volumes between neighboring positions.**

To further test the usefulness of the model, simulated diffraction patterns from amorphous crystalline composites were tested. For these samples, the final nanocrystal diameters were ~2.5 nm and ~1.7 nm for the models that were created with initial radii of 4× and 3× the lattice parameter of Cu. Because the quench rate in MD is far above experimental values, the resulting glassy matrices can be less structurally relaxed and can contain excess free volume. The frozen-nanocrystal approach imposes the size and placement of crystalline regions without explicit nucleation and growth, so the resulting interfacial strain fields are physically plausible but do not correspond to a specific processing pathway. As a result, these composites may exhibit SSRO that differs from experimentally annealed glasses, but they provide a controlled set of amorphous–crystalline environments for testing the model's transferability. Figure 9 shows the predicted versus true disorder parameter from profiles taken across these samples, with one pathway crossing the embedded nanocrystals (Figure 9(a)-(c)) and another crossing the amorphous-amorphous interface in the composites (Figure 9(d)-(f)). The predictions generally closely align with the true $d_i$ in all scans. When crossing the nanocrystals, the disorder parameter drops precipitously, although not exactly to zero. This is due to the regions sampled for simulated diffraction having a height of 20 Å, which exceeds the diameter of the nanocrystals. The effect is more obvious for the smaller nanocrystals in Figure 9(c). When moving across the amorphous-crystalline interfaces, disorder parameter drops as this feature is more ordered than the amorphous matrix, yet only reaches intermediate values. Disorder parameters that are similar to those measured in the ACTRs in Figure 4(b) are observed. The errors between predicted and true values generally are larger for the sample with smaller nanocrystals, likely due to the fact that interface curvature is more pronounced in these specimens. The ability to provide accurate descriptors of embedded nanocrystals and their interfaces in amorphous-crystalline composites is missing from experiments, yet has been a focus of computational work. For example, Brandl et al. [70] investigated these features in layered amorphous-crystalline composites with MD simulations, finding that significant distortion can occur which can stabilize and trap dislocations at the interfacial plane. The ability to quantify the structure of the amorphous-crystalline interfaces will also enrich combined computational experimental studies such as those pioneered by Abdelmawla et al. [71]. The ability to quantify and spatially map local SSRO in amorphous materials with NBED patterns could be used in the future to test hypotheses about amorphous materials, such as the paracrystalline model discussed in the work of Voyles et al. [72], by enabling quantitative tests of motif connectivity and spatial heterogeneity. While diffraction-derived approaches commonly used in NBED and 4D-STEM, such as radial fitting of amorphous halos or comparing Bragg-like intensity to diffuse scattering, would be expected to show similar spatial trends to the $d_i$ maps here, our workflow provides a simulation-native and directly interpretable SSRO descriptor that enables direct comparison with atomistic modeling. Symmetry-specific approaches such as correlation symmetry analysis can be complementary to relate disorder magnitude to specific local symmetry signatures in future work [73].

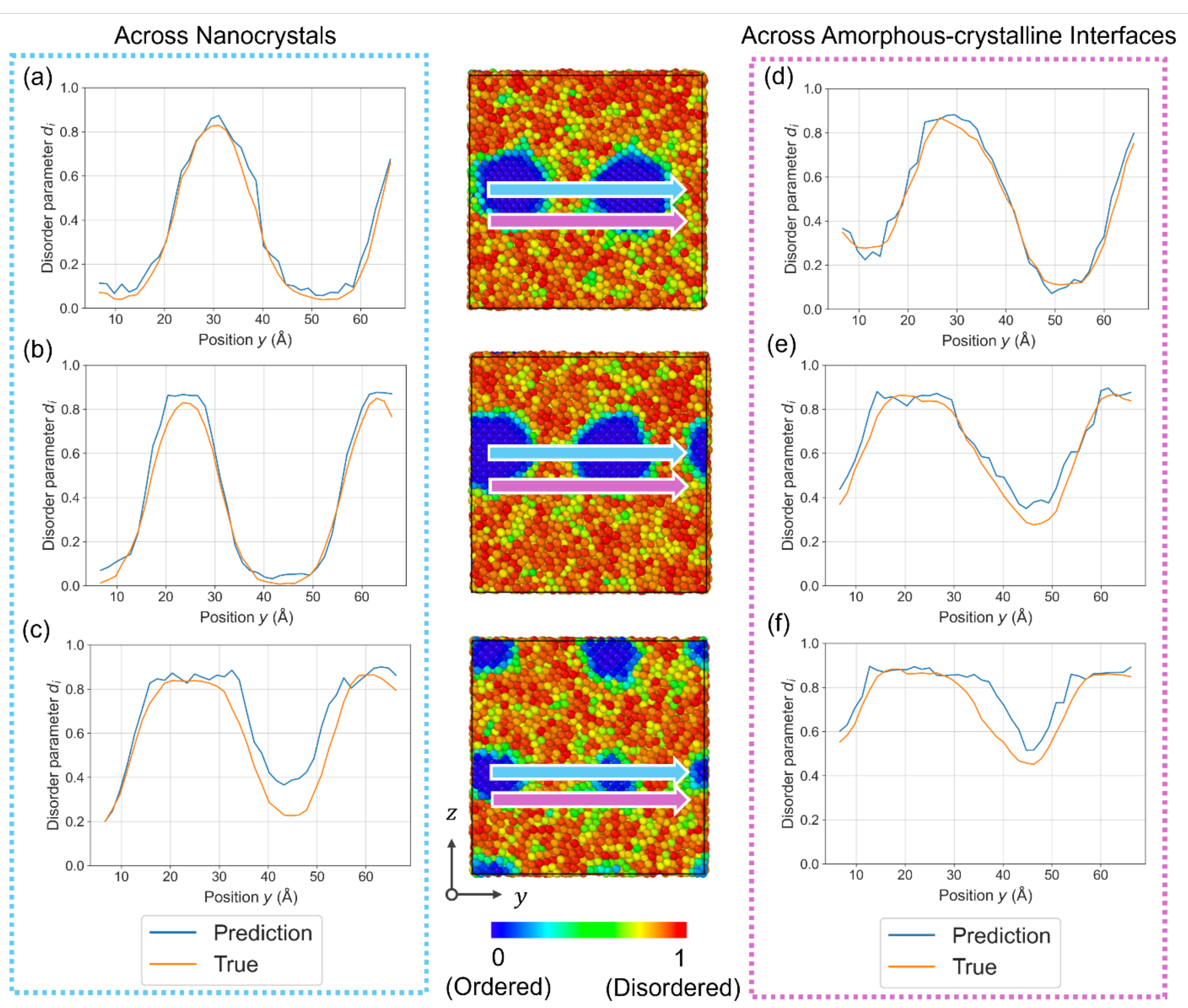

**Figure 9. Profiles of the predicted and true $d_i$ for amorphous-crystalline composites containing nanocrystals of varying sizes: (a, d) radius of 4 lattice parameters with varied orientations, (b, e) radius of 4 lattice parameters with aligned orientations, and (c, f) radius of 3 lattice parameters with aligned orientations. The left column (a, b, c) shows scans passing through the nanocrystals, while the right column (d, e, f) shows scans passing through the amorphous-crystalline interface. Arrows indicate the positions of the profiles.**

To validate the model's ability to extract disorder parameter from experimental NBED images, the model was tested on published data on metallic glasses. Figure 10(a) shows representative images from the three datasets used here. The first two come from an arc melted $Zr_{50}Cu_{50}$ metallic glass, where Islam et al. [74] were interested in the classification of medium-range order with different symmetries. As such, this data is subdivided into patterns with either 2-fold or 6-fold symmetry. It is worth noting that the presence of symmetry does not necessarily mean a structure is ordered. For example, Hirata et al. [41] showed clear symmetries in distorted icosahedral packing motifs that were generally very disordered. The third dataset comes from a $Zr_{50}Cu_{40}Al_{10}$ metallic glass that was investigated with NBED while being relaxed near its glass transition temperature by Nakazawa et al. [75]. Figure 10(b) presents the predicted disorder parameter for each dataset, with all predictions being relatively disordered and appropriate for a glassy structure. The values for the two $Zr_{50}Cu_{50}$ sets are similar and have relatively high values, yet the mean values of the two predictions are distinguishable and differ by ~0.01. Icosahedral motifs have both 2- and 6-fold symmetries and thus a high value is reasonable. The $Zr_{50}Cu_{40}Al_{10}$ sample that is undergoing relaxation has lower disorder parameters, consistent with the idea that the relaxation process allows the atomic structure to find more ordered configurations. Although there are only limited statistics in this dataset (six NBED patterns from three distinct times), longer relaxation times result in lower values of the disorder parameter. While the ground truths cannot be known here, it is promising that the values measured here (1) fall within the expected ranges for amorphous solids and (2) are distinguishable for different glasses, comparing the $Zr_{50}Cu_{50}$ and $Zr_{50}Cu_{40}Al_{10}$ alloys. Future investigations with NBED, particularly those involving known ground truths which display a gradient from ordered crystalline to disordered states, such amorphous grain boundary complexions, would greatly enhance model validation. In addition, it is important to note that the current model was only trained on Cu-Zr glasses structures and does not account for variations in the scattering factors among different atomic species. Thus, further research, such as the construction of a comprehensive database, is crucial for systems with substantial atomic size mismatches.

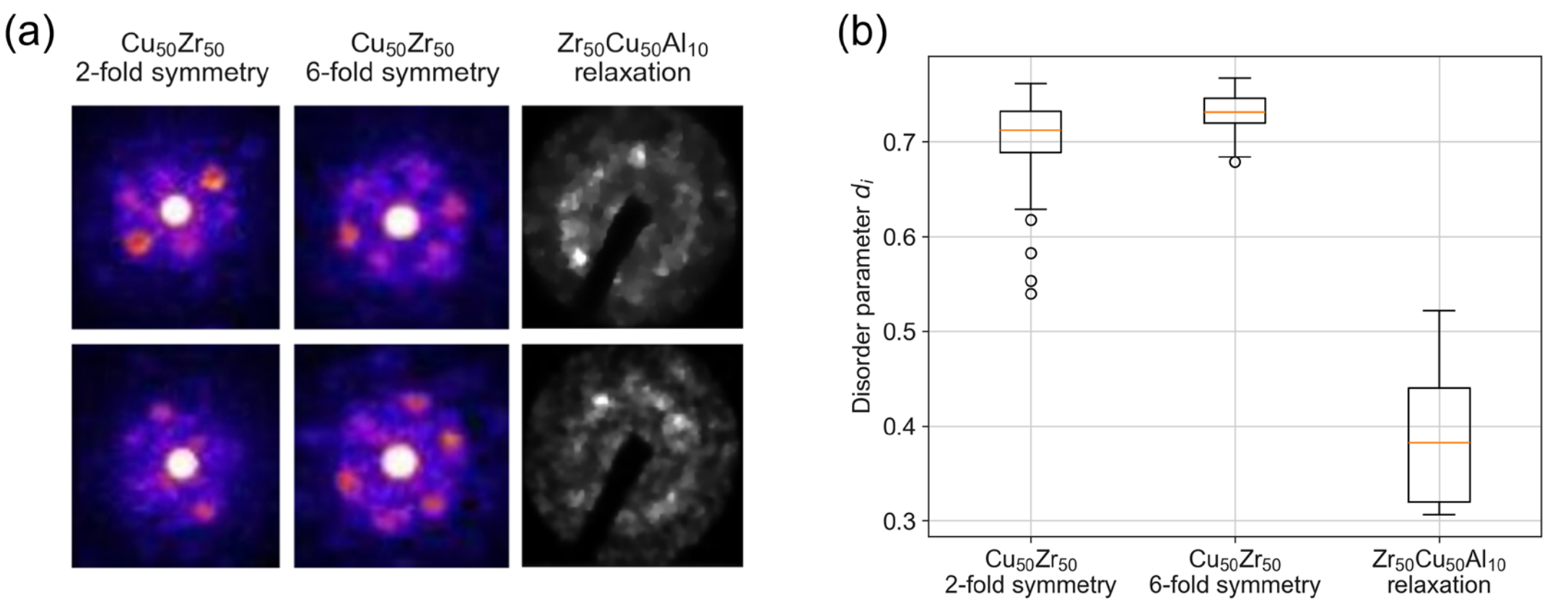

**Figure 10. NBED patterns from (a) a $Cu_{50}Zr_{50}$ metallic glass with 2-fold and 6-fold symmetries, from Islam et al. [74] and reproduced with permission of Oxford University Press, and a $Zr_{50}Cu_{40}Al_{10}$ undergoing relaxation near its glass transition temperature, from Nakazawa et al. [75] and reproduced with permission of Springer Nature. Images were denoised using total variation denoising implemented in scikit-image package [76,77] with a weight of 0.1 and a stop criterion of either $\epsilon = 0.0002$ or 200 epochs. Prior to model input, images were resized to 133 × 133 and pixel intensities were normalized to the [0, 1] range to match the model input dimensions. The complete set of processed experimental images is provided in Supplementary Note 4. (b) Predicted disordered parameters from the three datasets, all being within reasonable values for an amorphous metal and with the sample undergoing relaxation exhibiting a less disordered structure. The boxes represent the first quartile to third quartile range, with a center line in orange representing the median. The whiskers extend 1.5 times the interquartile range from the boxes. Data points marked as circles denote the outliers lying outside of the end of the whiskers.**

## 4. Summary and Conclusions

An ML-driven analysis approach was created to quantify SSRO in amorphous solids using NBED and 4D-STEM data. By moving beyond analysis limited to discrete atomic motifs, our regression-based model uses the disorder parameter $d_i$ to provide a continuous and interpretable metric of SSRO. Through transfer learning with ResNet-18, the model accurately predicts disorder parameter profiles using varied diffraction volume geometries, including cylinders and ellipsoids that compare favorably to experimental reality. In addition, the dependence on the geometry of the diffraction volume was probed to identify optimal combinations of radius and height, providing guidance for future experiments. The model was validated on unseen amorphous grain boundary complexions and amorphous-crystalline composite structure, with excellent agreement between predictions and true values, demonstrating that the framework is transferable. When applied to external experimental NBED datasets, the model correctly identified structural relaxation trends and symmetry-related disorder levels, confirming its potential for experimental applications. Future research will focus on expanding the training database to include a broader range of alloy systems with diverse atomic size mismatches and scattering factors. The analysis framework developed here will enable experimental mapping of SSRO in amorphous solids, ultimately aiding in the rational design of amorphous alloys with tailored performance.

**Data Availability**

Codes related to the model, data processing, and analysis used in this work are available in a GitHub repository at https://github.com/NMMLab/nbed-ssro.

**Acknowledgements**

This research was supported by the U.S. Department of Energy, Office of Science, Basic Energy Sciences, under Award No. DE-SC0025195.

## Supplementary Materials

### Supplementary Note 1. Voronoi Edge Length Threshold

Voronoi tessellation is sensitive to small perturbations in atomic positions, which can generate extremely small faces and edges associated with distant atoms that should not be considered part of the nearest-neighbor environment. To reduce this sensitivity, an edge-length cutoff is commonly applied so that faces bounded by unrealistically short edges are excluded from the Voronoi index calculation.

To select an appropriate cutoff, we computed the distribution of Voronoi edge lengths for atoms in the Cu–Zr grain boundary complexion. As shown in Figure S1(a), most edges form a broad primary distribution centered near ∼1.5 Å, while a secondary peak appears at very small edge lengths. The zoomed-in view in Figure S1(b) shows that these peaks are below ~0.5 Å. These extremely short edges arise from nearly degenerate geometric constructions and are not physically meaningful descriptors of the local packing environment. An example is shown in Figure S1(c), where the tessellation produces a degenerate face associated with an atom that is far from the center atom and would not be expected to participate in the local coordination shell. In this case, the distance between the center Zr atom and the atom indicated by the arrow is 3.75 Å, which is substantially larger than the typical distances to the nearest neighbors (average ∼2.72 Å) and longer than a typical Zr-Zr nearest-neighbor distance. These degeneracies can also lead to incorrect Voronoi indices even for otherwise well-ordered structures. Figure S1(d) shows a nominally FCC local environment that is assigned an incorrect index of $\langle 0, 6, 0, 8 \rangle$ when very small faces are retained, because two additional, distant atoms are spuriously counted as neighbors. After applying an edge-length cutoff, the same central atom yields the correct FCC index $\langle 0, 12, 0, 0 \rangle$, as shown in Figure S1(e).

Based on Figure S1(a), an edge-length cutoff of 0.6 Å was selected, which removes the degenerate, short edges while preserving the physically meaningful portion of the edge length distribution. This cutoff was used consistently for all Voronoi index calculations reported in the manuscript.

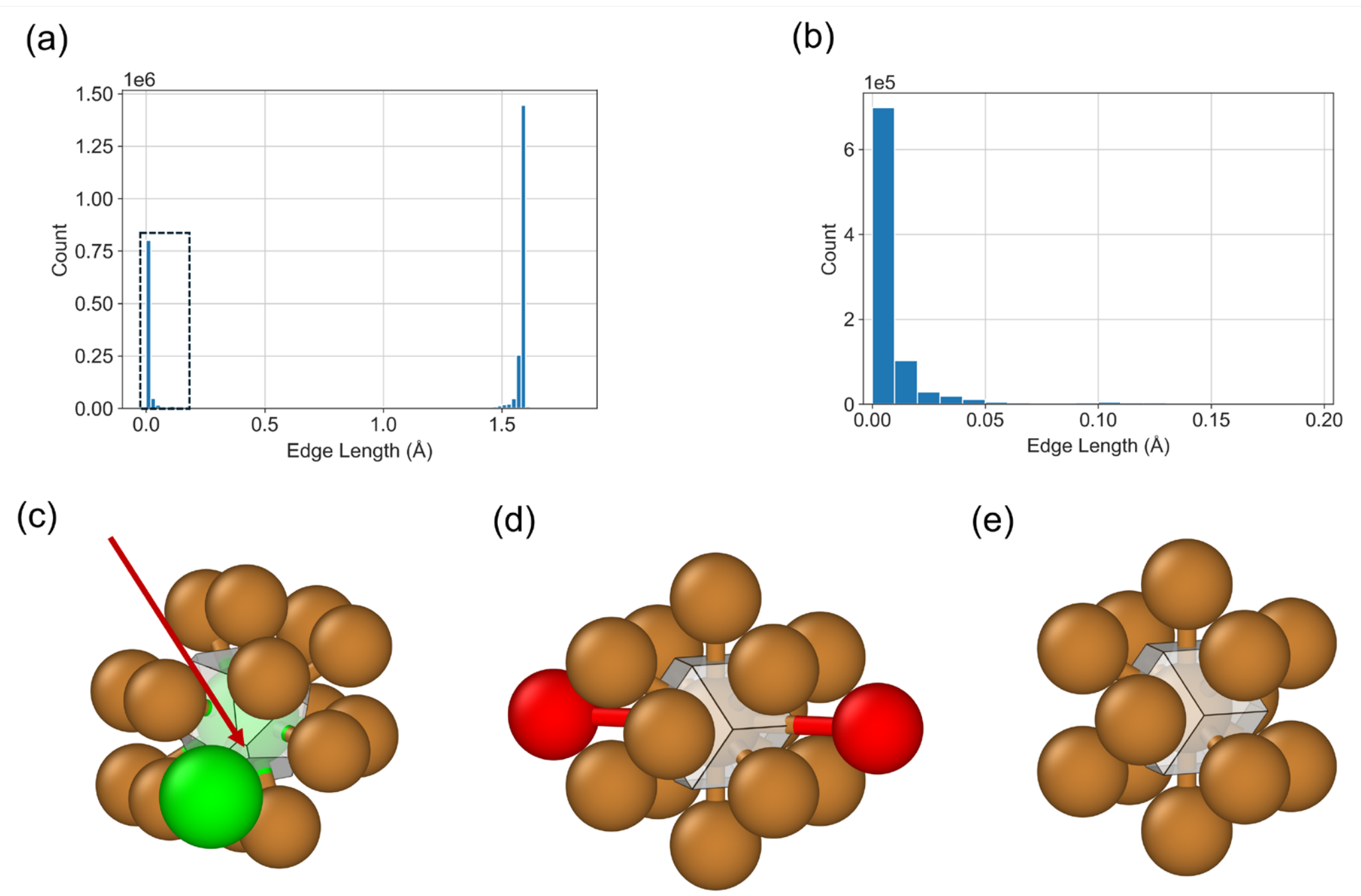

**Figure S1. (a) Distribution of distribution of edge length for Voronoi polyhedra in the Cu-Zr grain boundary complexion with no edge length cutoff applied and (b) a zoomed-in view of the short edge length regime. (c) Example of Voronoi polyhedron containing a degenerate face denoted by the red arrow. (d) Degenerate FCC Voronoi polyhedron assigned an incorrect Voronoi index $< 0, 6, 0, 8 >$. The two spurious neighbors are shown in red. (e) Correctly constructed FCC Voronoi polyhedron for the same central atom after applying a cutoff, yielding the FCC index $< 0, 12, 0, 0 >$. In the polyhedron examples, the gray surface mesh enclosing the center atom represent the faces of the Voronoi polyhedron, and the sticks represent the "bonds" between the center atom and its neighbors. These faces are perpendicular to the bonds and are equidistant to the associated atoms. The atomic radii are scaled by a factor of 0.75 to clearly show the bonds, edges and faces, and as a result, the bonds might look longer than they really are.**

## Supplementary Note 2. ACTR Identification Criterion

ACTR atoms in Figure 4(b) were identified using a distance-to-interface criterion based on the minimum distance to the crystalline-amorphous interface. This is done by first identifying the crystalline-amorphous interface from the crystalline atoms classified by CNA and then labeled amorphous atoms within 5 Å of this interface (measured as the closest distance in *X*-direction to the interface surface) as belonging to the ACTR and amorphous atoms farther than 5 Å were labeled as complexion interior. The value of 5 Å was adopted from prior work by Garg and Rupert [1], where it was justified based on the spatial distribution of SSRO, and is consistent with our own analysis of individual slices used to create Figure 2. This 5 Å cutoff is a local proximity criterion used for partitioning atoms near a rough interface.

## Supplementary Note 3. Voronoi Polyhedra versus Disorder Parameter $d_i$

Figure S2 shows representative Voronoi polyhedra spanning the full range from highly ordered ($d_i$ = 0.0) to highly disordered ($d_i$ = 1.0). Aside from the ideal FCC case (<0, 12, 0, 0>), the Voronoi index alone does not provide a monotonic or unique indication of disorder.

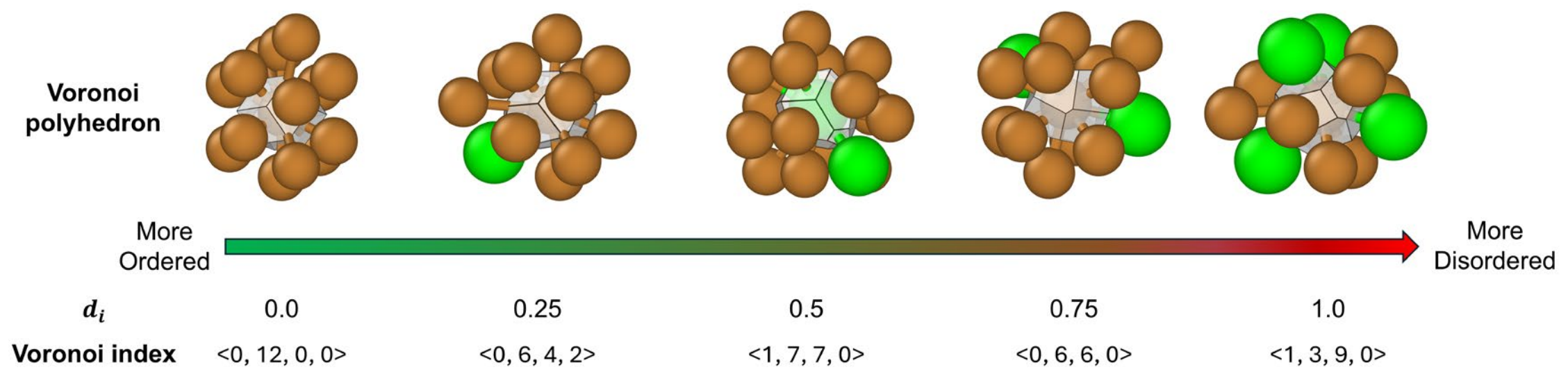

**Figure S2. Voronoi polyhedra examples with their Voronoi index and disorder parameter $d_i$ values.**

## Supplementary Note 4. Processed Experimental Images Used in Model Testing

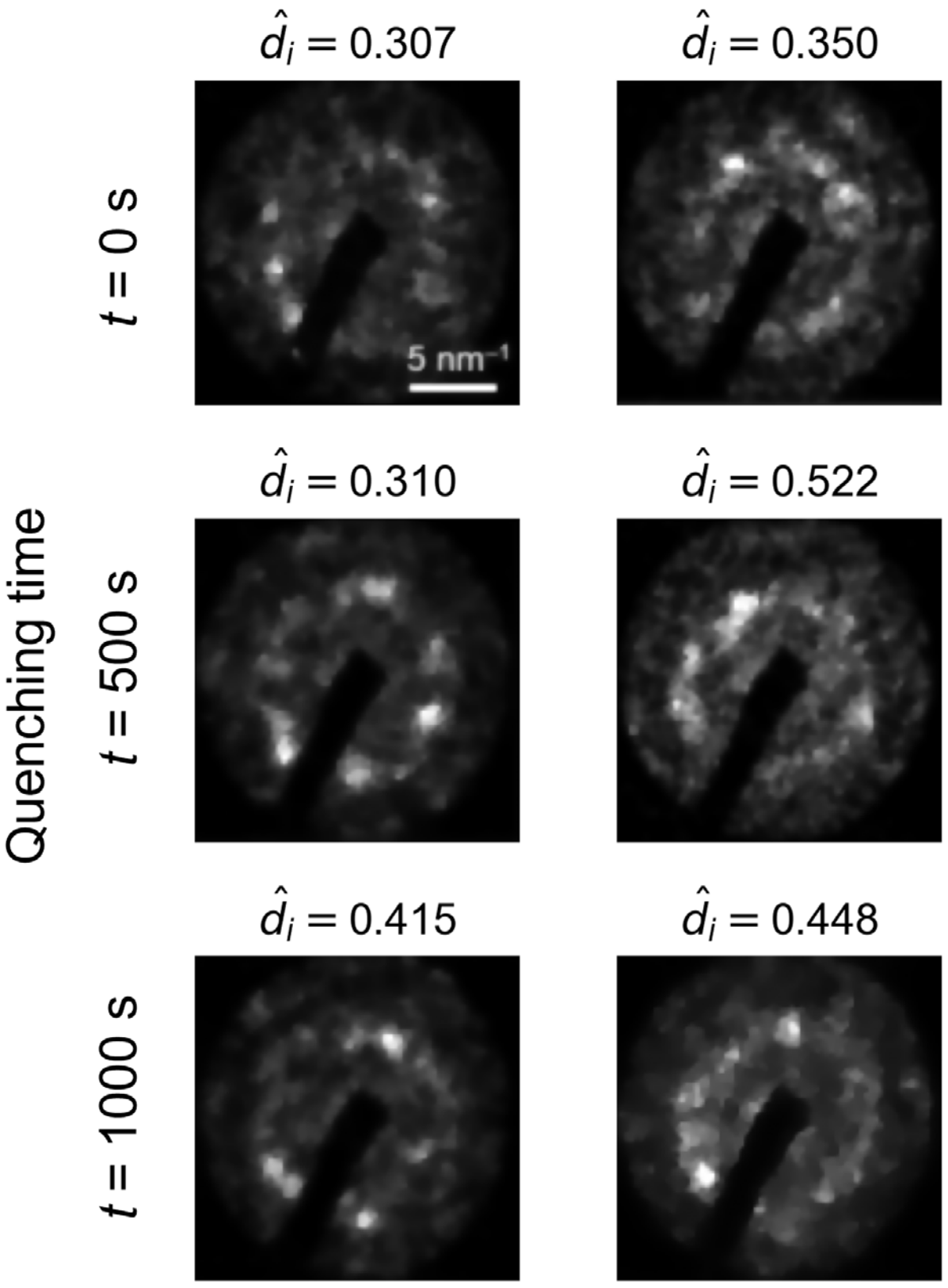

**Figure S3. Experimental NBED patterns from a $Zr_{50}Cu_{40}Al_{10}$ glass undergoing relaxation near its glass transition temperature, from Nakazawa et al. [2] and reproduced with permission of Springer Nature. Images were denoised using total variation denoising implemented in scikit-image package [3,4] with a weight of 0.1 and a stop criterion of either $\epsilon = 0.0002$ or 200 epochs. Prior to model input, images were resized to 133 × 133 and pixel intensities were normalized to the [0, 1] range to match the model input dimensions. The predicted $d_i$ values are shown at the top of each pattern.**

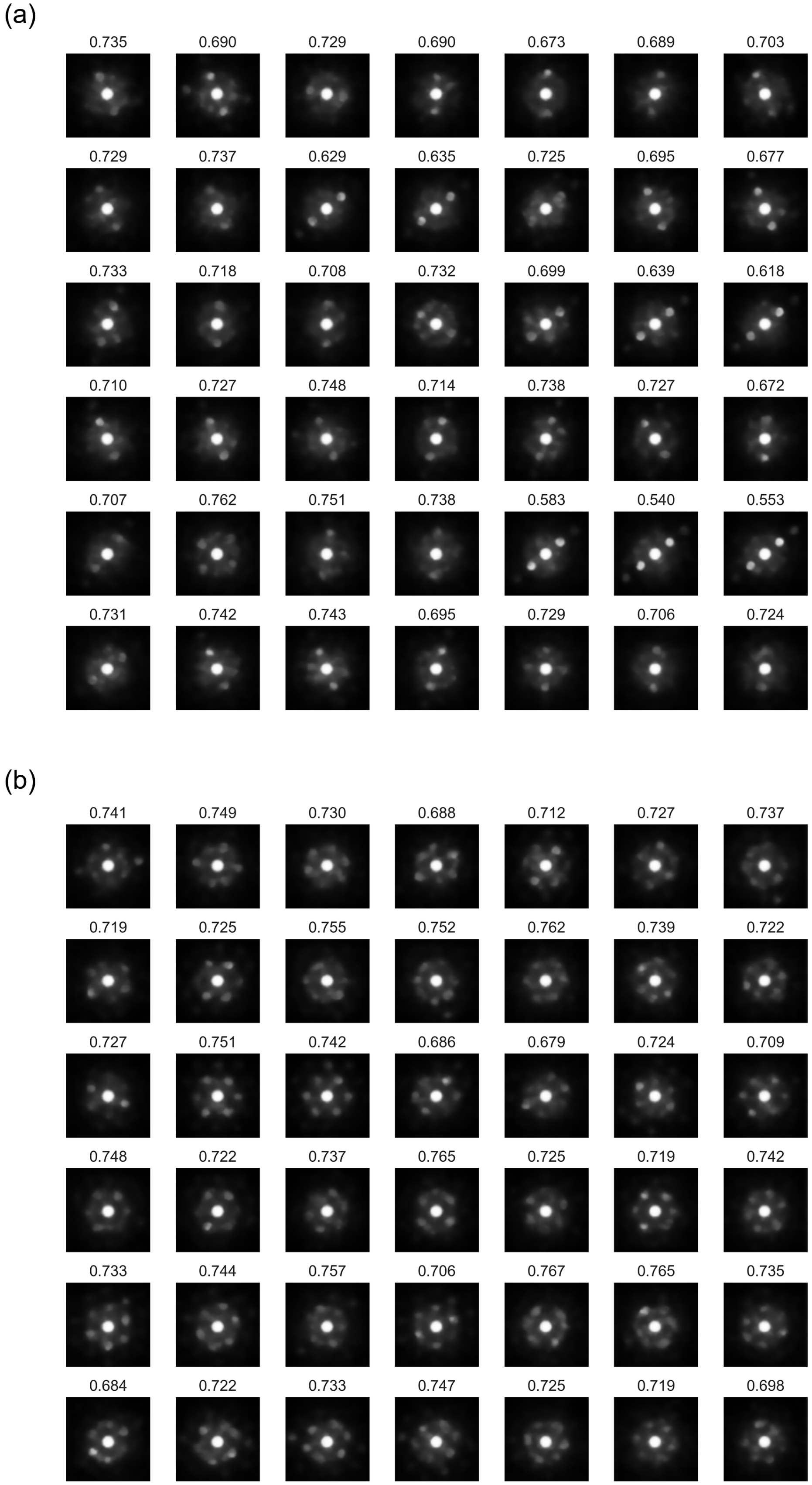

**Figure S4. NBED patterns from a $Cu_{50}Zr_{50}$ metallic glass with (a) 2-fold and (b) 6-fold symmetries, from Islam et al. [5] and reproduced with permission of Oxford University Press, Images were denoised using total variation denoising implemented in scikit-image package [3,4] with a weight of 0.1 and a stop criterion of either $\epsilon = 0.0002$ or 200 epochs. Prior to model input, images were resized to 133 × 133 and pixel intensities were normalized to the [0, 1] range to match the model input dimensions. The predicted $d_i$ values are shown at the top of each pattern.**